\documentclass[11pt]{article} 

\usepackage{amsmath}
\usepackage{amssymb}     
\usepackage{latexsym}   

\input diagrams.tex

\newcommand{\dda}{\mathord{\mbox{\makebox[0pt][l]{\raisebox{-.4ex}
                           {$\downarrow$}}$\downarrow$}}}
\newcommand{\dua}{\mathord{\mbox{\makebox[0pt][l]{\raisebox{.4ex}
                           {$\uparrow$}}$\uparrow$}}}
\def\endproof{$\ \ \Box$} 

\newcommand{\bq}{\begin{quote}}
\newcommand{\eq}{\end{quote}}

\newcommand{\Cl}{\mathrm{Cl}}
\newcommand{\reals}{{\mathbb R}}

\newcommand{\UX}{{\mathbf U}\,\!{\mathit X}}

\newcommand{\rat}{{\mathbb Q}}

\newtheorem{Th}{Theorem}[section]
\newtheorem{theorem}[Th]{Theorem}
\newtheorem{proposition}[Th]{Proposition}     
\newtheorem{lemma}[Th]{Lemma}

\newtheorem{definition}[Th]{Definition} 
\newtheorem{example}[Th]{Example}

\begin{document}

\title{Spacetime topology from causality}

\author{Keye Martin \\ \\
{\small Department of Mathematics}\\
{\small Tulane University}\\
{\small New Orleans, LA 70118}\\
{\small United States of America}\\
{\small \texttt{martin@math.tulane.edu}}  \\ \\
Prakash Panangaden\\ \\
{\small School of Computer Science}\\
{\small McGill University}\\
{\small Montreal, Quebec H3A 2A7}\\
{\small Canada}\\
{\small \texttt{prakash@cs.mcgill.ca} }\\
}

\date{}
\maketitle

\begin{abstract} We prove that a globally
hyperbolic spacetime with its causality relation
is a bicontinuous poset whose interval topology
is the manifold topology. This provides an
abstract mathematical setting in which one
can study causality independent of geometry
and differentiable structure.
\end{abstract}

\section{Introduction}

It has been known for some time that
the topology of spacetime could be characterized
purely in terms of causality -- but what is causality?
In this paper, we prove that the causality
relation is much more than a relation -- it
turns a globally hyperbolic spacetime into what is known as a \em bicontinuous poset. \em
The order on a bicontinuous poset allows 
one to define an intrinsic topology called \em the interval topology. \em
On a globally hyperbolic spacetime,
the interval topology is the manifold topology.

The importance of these results and ideas
is that they suggest an abstract formulation
of causality -- a setting where one can study
causality independently of geometry
and differentiable structure. %In a forthcoming
%paper, some of its immediate benefits will be revealed.

\section{Domains, continuous posets and topology}

A \em poset \em is a partially ordered set, i.e., a set together
with a reflexive, antisymmetric and transitive relation.

\begin{definition}\em Let $(P,\sqsubseteq)$ be a partially ordered set.
  A nonempty subset $S\subseteq P$ is \em directed \em if $(\forall
  x,y\in S)(\exists z\in S)\:x,y\sqsubseteq z$. The \em supremum \em
  of $S\subseteq P$ is the least of all its upper bounds
  provided it exists. This is written $\bigsqcup S$.
\end{definition}
These ideas have duals
that will be important to us: A nonempty $S\subseteq P$ is \em filtered \em if $(\forall
  x,y\in S)(\exists z\in S)\:z\sqsubseteq x,y$. The \em infimum \em $\bigwedge S$
  of $S\subseteq P$ is the greatest of all its lower bounds
  provided it exists.

\begin{definition}\em For a subset $X$ of a poset $P$, set
\[\uparrow\!\!X:=\{y\in P:(\exists x\in X)\,x\sqsubseteq y\}\ \ \&\
\downarrow\!\!X:=\{y\in P:(\exists x\in X)\,y\sqsubseteq x\}.\] We
write $\uparrow\!x=\,\uparrow\!\{x\}$ and
$\downarrow\!x=\,\downarrow\!\{x\}$ for elements $x\in X$.
\end{definition}

A partial order allows for the derivation of
several intrinsically defined topologies. 
Here is our first example.

\begin{definition}\em  A subset $U$ of a poset $P$ is \em Scott open \em if
\begin{enumerate}
\item[(i)] $U$ is an upper set: $x\in U\ \&\ x\sqsubseteq y\Rightarrow
  y\in U$, and \item[(ii)] $U$ is inaccessible by directed suprema:
  For every directed $S\subseteq P$ with a supremum,
\[\bigsqcup S\in U\Rightarrow S\cap U\neq\emptyset.\]
\end{enumerate}
The collection of all Scott open sets on $P$ is called the \em Scott
topology. \em
\end{definition}

\begin{definition}\em
A \em dcpo \em is a poset in which every directed subset has a
  supremum. The \em least element \em in a poset,
  when it exists, is the unique element $\bot$ with $\bot\sqsubseteq x$
for all $x$.
\end{definition}

The set of \em maximal elements \em in a dcpo $D$ is
\[\max(D):=\{x\in D :\ \uparrow\!\!x=\{x\}\}.\]
Each element in a dcpo has a maximal
element above it.

\begin{definition}\em
  For elements $x,y$ of a poset, write $x\ll y$ iff for all directed
  sets $S$ with a supremum,
\[y\sqsubseteq\bigsqcup S\Rightarrow (\exists s\in S)\:x\sqsubseteq s.\]
We set $\dda x=\{a\in D:a\ll x\}$ and $\dua x=\{a\in D:x\ll a\}$.
\end{definition}
For the symbol ``$\ll$,'' read ``approximates.'' 

\begin{definition}\em
  A \em basis \em for a poset $D$ is a subset $B$ such that $B\cap\dda x$
  contains a directed set with supremum $x$ for all $x\in D$.  A poset is
  \em continuous \em if it has a basis. A poset is $\omega$-\em continuous \em
  if it has a countable basis.
\end{definition}

Continuous posets have an important property, they are \em interpolative. \em

\begin{proposition} If $x\ll y$ in a continuous poset $P$, then
there is $z\in P$ with $x\ll z\ll y$.
\end{proposition}

This enables a clear description of the Scott topology,

\begin{theorem}
  The collection $\{\dua x:x\in D\}$ is a basis for the Scott topology
  on a continuous poset.
\end{theorem}

And also helps us give a clear definition of the \em Lawson topology. \em

\begin{definition}\em The \em Lawson topology \em on a continuous poset $P$
has as a basis all sets of the form $\dua x\setminus\!\!\uparrow\!\!F$,
for $F\subseteq P$ finite.
\end{definition}
The next idea, as far as we know, is new.
\begin{definition}\em A continuous poset $P$ is \em bicontinuous \em if
\begin{itemize}
\item For all $x,y\in P$, $x\ll y$ iff for all filtered $S\subseteq P$
with an infimum,
\[\bigwedge S\sqsubseteq x\Rightarrow (\exists s\in S)\,s\sqsubseteq y, \]
and
\item For each $x\in P$, the set $\dua x$ is filtered with infimum $x$.
\end{itemize}
\end{definition}

\begin{example}\em $\reals$, $\rat$ are bicontinuous.
\end{example}

\begin{definition}\em On a bicontinuous poset $P$, sets of the form
\[(a,b):=\{x\in P:a\ll x\ll b\}\]
form a basis for a topology called \em the interval topology. \em
\end{definition}
The proof uses interpolation and bicontinuity. A 
bicontinuous poset $P$ has $\dua x\neq\emptyset$ for each $x$,
so it is rarely a dcpo.
Later we will see that
on a bicontinuous poset, the Lawson topology
is contained in the interval topology (causal simplicity),
the interval topology is Hausdorff (strong causality),
and $\leq$ is a closed subset of $P^2$.

\begin{definition}\em 
A \em continuous dcpo \em is a continuous poset which is also a dcpo.
A \em domain \em is a continuous dcpo.
\end{definition}
We now consider an example of a domain 
that will be used later in proofs.

\begin{example}\em Let $X$ be a locally compact Hausdorff space. Its \em upper space \em
\[\UX=\{\emptyset\neq K\subseteq X:K\mbox{ is compact}\}\]
ordered under reverse inclusion
\[A\sqsubseteq B\Leftrightarrow B\subseteq A\]
is a continuous dcpo: 
\begin{itemize}
\item For directed $S\subseteq\UX$, $\bigsqcup S=\bigcap S.$ 
\item For all $K,L\in \UX$, $K\ll L\Leftrightarrow L\subseteq\mbox{int}(K)$.
\item $\UX$ is $\omega$-continuous iff $X$ has a countable basis.
\end{itemize}
It is interesting here that the space $X$ can be recovered
from $\UX$ in a purely order theoretic manner:
\[X\simeq\max(\UX)=\{\{x\}:x\in X\}\]
where $\max(\UX)$ carries the relative Scott topology it 
inherits as a subset of $\UX.$ Several constructions
of this type are known.
\end{example}

\section{The causal structure of spacetime}

A \em manifold \em $\mathcal{M}$ is a locally Euclidean Hausdorff space
that is connected and has a countable basis. 
A connected Hausdorff manifold is paracompact iff it has a countable basis. 
A \em Lorentz metric \em on a manifold is
a symmetric, nondegenerate tensor field of type $(0,2)$
whose signature is $(- + + +)$.

\begin{definition}\em A \em spacetime \em is a real four-dimensional
smooth manifold $\mathcal{M}$ with a Lorentz metric $g_{ab}$.
\end{definition}

Let $(\mathcal{M},g_{ab})$ be a time orientable spacetime.
Let $\Pi^+_\leq$ denote the future directed
causal curves, and $\Pi^+_{<}$ denote
the future directed time-like curves.

\begin{definition}\em For $p\in \mathcal{M}$, 
\[I^+(p):=\{q\in\mathcal{M}:(\exists\pi\in\Pi^+_{<})\,\pi(0)=p, \pi(1)=q\}\]
and
\[J^+(p):=\{q\in\mathcal{M}:(\exists\pi\in\Pi^+_{\leq})\,\pi(0)=p, \pi(1)=q\}\]
Similarly, we define $I^-(p)$ and $J^-(p)$.
\end{definition}
We write the relation $J^+$ as
\[p\sqsubseteq q\equiv q\in J^+(p).\]
The following properties from~\cite{he} are very useful: 
\begin{proposition} 
\label{page183}
Let $p,q,r\in\mathcal{M}$. Then
\begin{enumerate}
\em\item[(i)]\em The sets $I^+(p)$ and $I^-(p)$ are open.
\em\item[(ii)]\em $p\sqsubseteq q$ and $r\in I^+(q)$ $\Rightarrow$ $r\in I^+(p)$
\em\item[(iii)]\em  $q\in I^+(p)$ and $q\sqsubseteq r$ $\Rightarrow$ $r\in I^+(p)$
\em\item[(iv)]\em $\Cl(I^+(p))=\Cl(J^+(p))$ and $\Cl(I^-(p))=\Cl(J^-(p))$.
\end{enumerate}
\end{proposition}

We always assume the chronology conditions that ensure $(\mathcal{M},\sqsubseteq)$
is a partially ordered set. 
We also assume \em strong causality \em
which can be characterized as follows~\cite{penrose}:

\begin{theorem}
\label{suspect} 
A spacetime $\mathcal{M}$ is strongly causal iff its Alexandroff topology
is Hausdorff iff its Alexandroff topology is the manifold topology.
\end{theorem}

The Alexandroff topology 
on a spacetime has ${\{I^+(p)\cap I^-(q):p,q\in\mathcal{M}\}}$ as a basis~\cite{penrose}.

\section{Global hyperbolicity}

Penrose has called \em globally hyperbolic \em spacetimes
``the physically reasonable spacetimes~\cite{wald}.'' In this section, 
$\mathcal{M}$ denotes a globally hyperbolic spacetime, and
we prove that $(\mathcal{M},\sqsubseteq)$ is a bicontinuous poset.

\begin{definition}\em A spacetime $\mathcal{M}$ is \em globally hyperbolic \em
if it is strongly causal and 
if $\uparrow\!\!a\ \cap\downarrow\!\!b$ is compact in the manifold topology,
for all $a,b\in\mathcal{M}$.
\end{definition}

\begin{lemma} 
\label{limitsAreSuprema}
If $(x_n)$ is a sequence in $\mathcal{M}$ with $x_n\sqsubseteq x$ for all n, then
\[\lim_{n\rightarrow\infty}x_n=x\ \Rightarrow\ \bigsqcup_{n\geq 1}x_n=x.\]
\end{lemma}
{\bf Proof}. Let $x_n\sqsubseteq y$. 
By global hyperbolicity, $\mathcal{M}$ is causally simple,
so the set $J^-(y)$ is closed. Since $x_n\in J^-(y)$, 
$x=\lim x_n \in J^-(y)$, and thus $x\sqsubseteq y$.
This proves $x=\bigsqcup x_n$.
\endproof

\begin{lemma} 
\label{seqapprox}
For any $x\in\mathcal{M}$, $I^-(x)$ contains
an increasing sequence with supremum $x$.
\end{lemma}
{\bf Proof}. Because $x\in\Cl(I^-(x))=J^-(x)$ but $x\not\in I^-(x)$,
$x$ is an \em accumulation point \em of $I^-(x)$,
so for every open set $V$ with $x\in V$, $V\cap I^-(x)\neq\emptyset$. 
Let $(U_n)$ be a countable basis for $x$, which
exists because $\mathcal{M}$ is locally Euclidean. Define
a sequence $(x_n)$ by first choosing
\[x_1\in U_1\cap I^-(x)\neq\emptyset\]
and then whenever
\[x_n\in U_n\cap I^-(x)\]
we choose 
\[x_{n+1}\in (U_n\cap I^+(x_n))\cap I^-(x)\neq\emptyset.\]
By definition, $(x_n)$ is increasing,
and since $(U_n)$ is a basis for $x$, $\lim x_n=x$.
By Lemma~\ref{limitsAreSuprema}, $\bigsqcup x_n=x$.
\endproof

\begin{proposition} 
\label{approx}
Let $\mathcal{M}$ be a globally hyperbolic spacetime. Then
\[x\ll y\Leftrightarrow y\in I^+(x)\]
for all $x,y\in\mathcal{M}$.
\end{proposition}
{\bf Proof}. Let $y\in I^+(x)$. Let $y\sqsubseteq \bigsqcup S$
with $S$ directed. By Prop.~\ref{page183}(iii),
\[y\in I^+(x)\ \&\ y\sqsubseteq \bigsqcup S\ \Rightarrow\ \bigsqcup S\in I^+(x)\]
Since $I^+(x)$ is manifold open and $\mathcal{M}$ is locally compact,
there is an open set $V\subseteq \mathcal{M}$ whose
closure $\Cl(V)$ is compact with $\bigsqcup S\in V\subseteq\Cl(V)\subseteq I^+(x)$.
Then, using approximation
on the upper space of $\mathcal{M}$, 
\[\Cl(V)\ll \left\{\bigsqcup S\right\}=\bigcap_{s\in S} [s,\bigsqcup S]\]
where the intersection on the right is
a filtered collection of nonempty compact sets by directedness
of $S$ and global hyperbolicity of $\mathcal{M}$. Thus,
for some $s\in S$, $[s,\bigsqcup S]\subseteq\Cl(V)\subseteq I^+(x)$,
and so $s\in I^+(x)$, which gives $x\sqsubseteq s$.
This proves $x\ll y$.

Now let $x\ll y$. By Lemma~\ref{seqapprox}, there
is an increasing sequence $(y_n)$ in $I^-(y)$
with $y=\bigsqcup y_n$. Then since $x\ll y$,
there is $n$ with $x\sqsubseteq y_n$. By Prop.~\ref{page183}(ii), 
\[x\sqsubseteq y_n\ \&\ y_n\in I^-(y)\ \Rightarrow\ x\in I^-(y)\]
which is to say that $y\in I^+(x)$.
\endproof

\begin{theorem} If $\mathcal{M}$ is globally hyperbolic, 
then $(\mathcal{M},\sqsubseteq)$ is a bicontinuous poset
with $\ll\ =I^+$ whose interval topology is the manifold topology.
\end{theorem}
{\bf Proof}. By combining Lemma~\ref{seqapprox} with Prop.~\ref{approx},
$\dda x$ contains an increasing sequence with supremum $x$, for each $x\in\mathcal{M}$.
Thus, $\mathcal{M}$ is a continuous poset. 

For the bicontinuity, Lemmas~\ref{limitsAreSuprema}, \ref{seqapprox} and
Prop.~\ref{approx} have ``duals'' which are obtained by replacing `increasing' by `decreasing', 
$I^+$ by $I^-$, $J^-$ by $J^+$, etc. For example, the dual of 
Lemma~\ref{seqapprox} is that $I^+$ contains a \em decreasing \em sequence
with \em infimum \em $x$. Using the duals of these two lemmas,
we then give an alternate characterization of $\ll$ in terms
of infima:
\[x\ll y\equiv (\forall S)\,\bigwedge S\sqsubseteq x\ \Rightarrow\ (\exists s\in S)\, s\sqsubseteq y\]
where we quantify over \em filtered \em subsets $S$ of $\mathcal{M}$. These
three facts then imply that $\dua x$ contains a decreasing
sequence with inf $x$. But because $\ll$ can be
phrased in terms of infima, $\dua x$ itself must be
filtered with inf $x$.

Finally, $\mathcal{M}$ is bicontinuous, so we know
it has an interval topology. Because $\ll=I^+$,
the interval topology is the one generated 
by the timelike causality relation, which
by strong causality is the manifold topology. 
\endproof\newline

Bicontinuity, as we have defined it here,
is really quite a special property, and some
of the nicest posets in the world are not bicontinuous. For example,
the powerset of the naturals $\mathcal{P}\omega$ is not bicontinuous,
because we can have $F\ll G$ for $G$ finite, and $F=\bigcap V_n$
where all the $V_n$ are infinite.

\section{Causal simplicity}

It is also worth pointing out before we close,
that causal simplicity also has a characterization
in order theoretic terms.

\begin{definition}\em A spacetime $\mathcal{M}$ is \em causally simple \em
if $J^+(x)$ and $J^-(x)$ are closed for all $x\in\mathcal{M}$.
\end{definition}

\begin{theorem} Let $\mathcal{M}$ be a spacetime and $(\mathcal{M},\sqsubseteq)$ a continuous poset with $\ll\,=I^+$. 
The following are equivalent:
\begin{enumerate}
\em\item[(i)]\em $\mathcal{M}$ is causally simple.
\em\item[(ii)]\em The Lawson topology on $\mathcal{M}$ is a subset of
the interval topology on $\mathcal{M}$.
\end{enumerate}
\end{theorem}
{\bf Proof} (i) $\Rightarrow$ (ii): We want to prove that
\[\{\dua x\:\cap\uparrow\!\!F:x\in\mathcal{M}\ \&\ F\subseteq\mathcal{M}\ \mbox{finite}\}\subseteq\mbox{int}_\mathcal{M}.\]
By strong causality of $\mathcal{M}$ and $\ll\,=I^+$,
$\mbox{int}_\mathcal{M}$ is the manifold topology,
and this is the crucial fact we need as follows.
First, $\dua x=I^+(x)$ is open in the manifold
topology and hence belongs to $\mbox{int}_\mathcal{M}$.
Second, $\uparrow\!x=J^+(x)$ is closed
in the manifold topology by causal simplicity,
so $\mathcal{M}\setminus\!\!\uparrow\!\!x$ belongs
to $\mbox{int}_\mathcal{M}$. Then $\mbox{int}_\mathcal{M}$
contains the basis for the Lawson topology
given above.

(ii) $\Rightarrow$ (i): First, since $(\mathcal{M},\sqsubseteq)$ is
continuous, its Lawson topology is Hausdorff, so $\mbox{int}_\mathcal{M}$
is Hausdorff since it contains the Lawson topology by assumption.
Since $\ll\,=I^+$,  $\mbox{int}_\mathcal{M}$ is the Alexandroff
topology, so Theorem~\ref{suspect} implies $\mathcal{M}$ is strongly
causal.

Now, Theorem~\ref{suspect} also tells
us that $\mbox{int}_\mathcal{M}$
is the manifold topology. Since
the manifold topology $\mbox{int}_\mathcal{M}$ 
contains the Lawson by assumption,
and since
\[J^+(x)=\ \uparrow\!\!x\ \ \mbox{and}\ \ J^-(x)=\ \downarrow\!\!x \]
are both Lawson closed (the second is Scott closed),
each is also closed in the manifold topology,
which means $\mathcal{M}$ is causally simple.
\endproof\newline

Note in the above proof that
we have assumed causally simplicity implies strong causality.
If we are wrong about this, then (i) above should be replaced
with `causal simplicity+strong causality'.

\section{Conclusion}

It seems there is something of a gap in the hierarchy of
causality conditions. One goes from global hyperbolicity
all the way down to causal simplicity. It might be
good to insert a new one in between these two. Some possible
candidates are to require $(\mathcal{M},\sqsubseteq)$ a continuous (bicontinuous) 
poset. All of these might provide generalizations of global hyperbolicity
for which a lot could probably be proved. Bicontinuity,
in particular, has the nice consequence that one
does not have to explicitly assume strong causality
as one does with global hyperbolicity. Is $\mathcal{M}$ bicontinuous
iff it is causally simple?

In the forthcoming~\cite{gr}, the fact that globally hyperbolic spacetimes
are bicontinuous will enable us to prove that spacetime can be reconstructed
from a countable dense set and the causality relation in a purely
order theoretic manner using techniques from an area known
as \em domain theory\em~\cite{scott:outline}.

\end{document}